\documentclass[11pt]{article}
\usepackage{amsmath, amssymb, amsthm, float, fullpage, graphicx, multirow,  multicol, parskip,   setspace, subcaption}
\usepackage{tikz}

\usepackage{hyperref}
\usepackage{natbib}
\usepackage{color}

\definecolor{PennRed}{RGB}{152, 30 50}
\definecolor{PennBlue}{RGB}{0, 44, 119}
\definecolor{PennGreen}{RGB}{94, 179,70}
\definecolor{PennViolet}{RGB}{141, 76, 145}
\definecolor{PennSkyBlue}{RGB}{14, 118, 188}
\definecolor{PennOrange}{RGB}{243, 117, 58}
\definecolor{PennBrightRed}{RGB}{223,82, 78}

\hypersetup{pdfborder = {0 0 0.5 [3 3]}, colorlinks = true, linkcolor = PennBrightRed, citecolor = PennSkyBlue}

\title{Protocol for an Observational Study on the Effects of Social Distancing on Influenza-Like Illness and COVID-19}
\author{Bo Zhang, Ting Ye, Siyu Heng, Michael Z. Levy, $\text{Dylan S. Small}^{\dagger}$\\
University of Pennsylvania}
\date{\today}

\begin{document}
\maketitle

\let\thefootnote\relax\footnotetext{$\dagger$ \textit{Address for Correspondence:} Dylan S. Small, Department of Statistics, The Wharton School, University of Pennsylvania, 3730 Walnut Street, Philadelphia, PA 19104, U.S.A. (E-mail: \textsf{dsmall@wharton.upenn.edu}).}

\def\C{\mathbb{C}}
\def\R{\mathbb{R}}
\def\Q{\mathbb{Q}}
\def\Z{\mathbb{Z}}
\def\N{\mathbb{N}}

\def\P{\mathbb{P}}
\def\E{\mathbb{E}}

\def\bR{\mathbf{R}}
\def\bZ{\mathbf{Z}}
\def\bX{\mathbf{X}}
\def\br{\mathbf{r}}
\def\bx{\mathbf{x}}

\begin{abstract}
\vspace{0.1in}
\noindent\textbf{Background:} The novel coronavirus disease (COVID-19) is a highly contagious respiratory disease that was first detected in Wuhan, China in December 2019, and has since spread around the globe, claiming more than 69,000 lives by the time this protocol is written. It has been widely acknowledged that the most effective public policy to mitigate the pandemic is \emph{social and physical distancing}: keeping at least six feet away from people, working from home, closing non-essential businesses, etc. There have been a lot of anecdotal evidences floating around, suggesting that social distancing has a causal effect on disease mitigation; however, few studies have investigated the effect of social distancing on disease mitigation in a transparent and statistically-sound manner. We propose to conduct a matched observational study using the U.S. Health Weather Map data by Kinsa Inc., the Social Distancing Scoreboard data by Unacast Inc., and the U.S. census data by the U.S. Census Bureau. 

\noindent\textbf{Methods and Analysis} We propose to perform an optimal non-bipartite matching to pair counties with similar observed covariates but vastly different average social distancing scores during the first week (March 16th through Match 22nd) of President's \emph{15 Days to Slow the Spread} campaign. We have produced a total of $302$ pairs of two U.S. counties with good covariate balance on a total of $16$ important variables. Our primary outcome will be the average observed illness collected by Kinsa Inc. two weeks after the intervention period. Although the observed illness does not directly measure COVID-19, it reflects a real-time aspect of the pandemic, and unlike confirmed cases, it is much less confounded by counties' testing capabilities. We also consider observed illness three weeks after the intervention period as a secondary outcome. We will test a proportional treatment effect using a randomization-based test with covariance adjustment and conduct a sensitivity analysis.


\end{abstract}

\newpage
\section{Background and Motivation}
\label{sec:background_motivation}

The 2019 novel coronavirus disease (COVID-19) is a highly infectious disease caused by a novel severe acute respiratory coronavirus (SARS-CoV-2) (\citealp{world2020novel}) that was first detected in Wuhan, China in December, 2019 (\citealp{huang2020clinical}; \citealp{zhu2020novel}) and has since quickly spread around the world. The ongoing outbreak has been declared by the World Health Organization (WHO) as a global pandemic (\citealp{globalpandemic}). The common symptoms include cough, fever, shortness of breath, diarrhea, sore throat, and muscle pain (\citealp{symptoms}), and around one out of every six patients progresses to develop difficulty breathing and become seriously ill (\citealp{symptoms}). As of April 5th, more than 1,270,000 cases have been confirmed in more than 200 countries and territories, resulting in over 69,300 deaths (\citealp{JHU}).

In the United States (U.S.), state and local governments have enacted social distancing policies designed to limit the spread of COVID-19 including banning mass gatherings, closing schools, promoting working at home, and closing non-essential businesses \citep{klein2020}. However, there has been considerable variation in how soon and how much social distancing jurisdictions have enacted.  The goal of this observational study is to assess whether heavier social distancing policies have slowed the spread of COVID-19 and influenza-like illness compared to lighter social distancing policies.

\section{Overview of the Study Design}

To study the treatment effect of social distancing on mitigating the spread of COVID-19, we first integrate several rich data sources, including the U.S. Health Weather Map data by Kinsa Inc., the Social Distancing Scoreboard data by Unacast Inc., U.S. census data by the United States Census Bureau, and the county-level public health assessment data by the County Health Rankings and Roadmaps program, using the Federal Information Processing Standards (FIPS) code unique to each county to produce a comprehensive dataset. We then construct a matched observational study out of this integrated dataset.

Matching is a nonparametric (i.e., without model assumptions) technique of adjusting for the observed covariates to control for potential bias and widely used in observational studies (\citealp{rubin1973matching}; \citealp{rosenbaum2002observational}). Our general idea is to pair counties that had a high social distancing score during the first week (March 16th through 22nd) of White House's \emph{15 Days to Slow the Spread} campaign with those that had a low social distancing score but are otherwise very similar, and compare observed influenza-like illness after two weeks (March 30th through April 5th, the primary outcome), and three weeks (April 6th through 12th, the secondary outcome).

The output of our statistical matching procedure is $302$ matched pairs of two counties with similar observed covariates, including population density, population composition, health care index, and observed illness during the treatment period (March 16th through 22nd). We will refer to the county with higher social distancing score in each matched pair as the \emph{treated} county, and the other the \emph{control} county. 

Treatment, primary outcome, and a full list of observed covariates to be controlled in the study are detailed in Section \ref{sec: treatment}, \ref{sec: outcome}, and \ref{sec: confounding vbs}.


\section{Definition of Heavier vs. Lighter Social Distancing Policies}
\label{sec: treatment}
As of April 5th, mobility data company Unacast Inc. measures the degree of social distancing on a daily basis for U.S. counties based on two metrics: 1) reduction in distance travelled, and 2) reduction in visits to non-essential venues, where the categorization of essential and non-essential venues are based on the guidelines issued by various state governments and policy makers. Each county is given a grade ($A$ through $F$) based on its reduction in distance traveled and another grade ($A$ through $F$) based on reduction in visits to non-essential venues. These two grades are then combined together to yield an overall social distancing score ($A$ through $F$). We transform the letter grade to the following $0-4$ scale: $A=4.0, A-=3.7, B=3.0, B-=2.7, C=2, C-=1.7, D=1, D-=0.7, F=0$.

\begin{figure}
    \centering
    \includegraphics[scale=0.6]{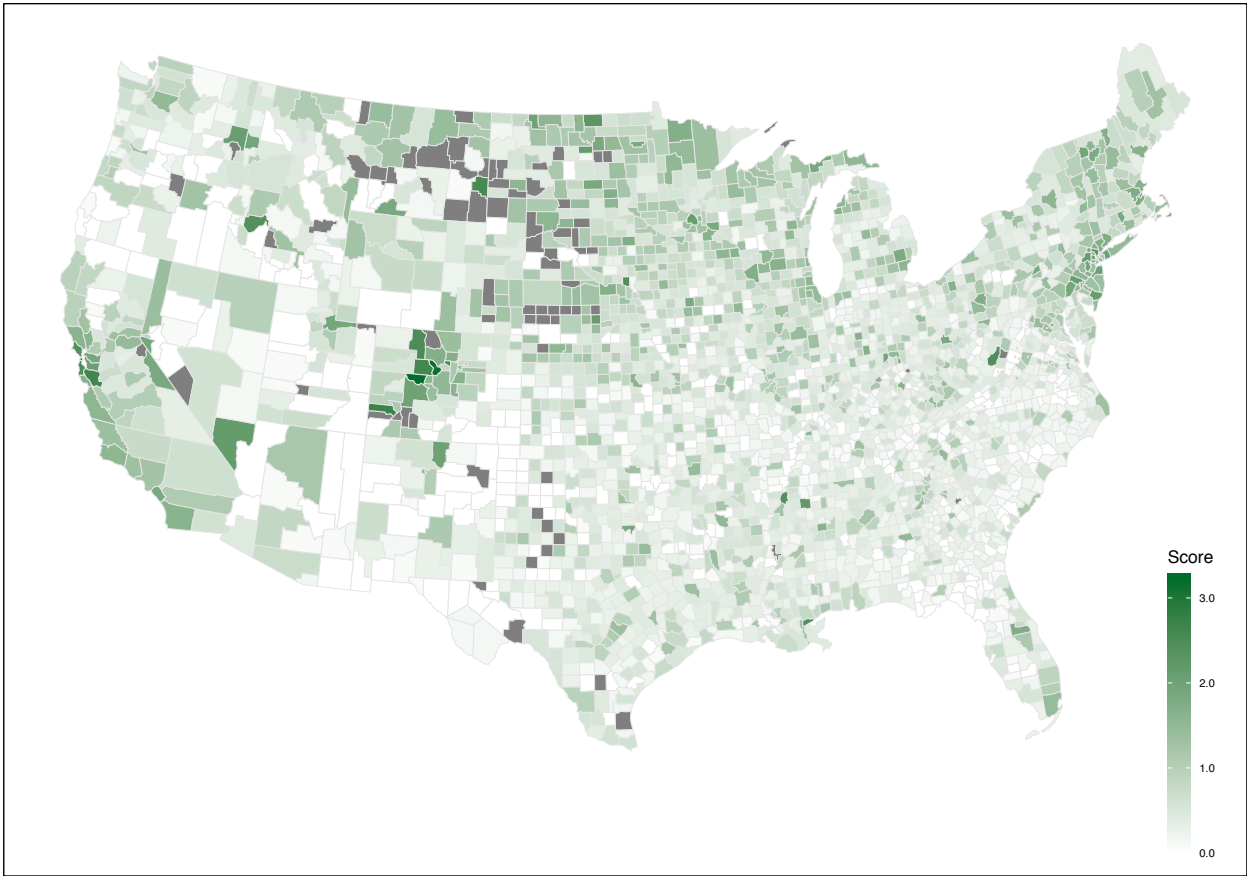}
    \caption{Visualization of the average overall social distancing score for every U.S. county (not including counties in Alaska and Hawaii). Gray represents missing data.}
    \label{fig:usmap_ATG}
\end{figure}

The intervention of interest is the average overall social distancing score during the first week (March 16th through March 22nd) of White House's first \emph{15 Days to Slow the Spread} campaign. Figure \ref{fig:usmap_ATG} visualizes this average social distancing score for every county during this time period.

\section{Outcomes}
\label{sec: outcome}
\begin{figure}
    \centering
    \includegraphics[scale = 0.8]{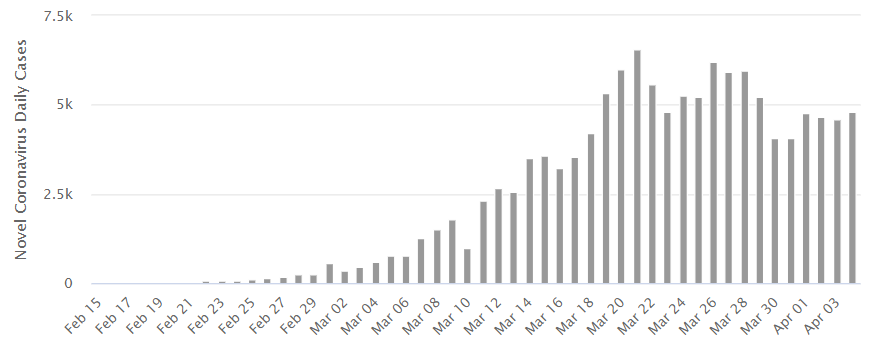}
    \caption{Daily News Cases in Italy}
    \label{fig: italy new cases}
\end{figure}

As of April 5th, there is no consensus as of when the effect of social distancing, if there is any, would emerge. Hypotheses are often drawn from countries that had lived through the pandemic curve. Figure \ref{fig: italy new cases} published by \hyperlink{https://www.worldometers.info/coronavirus/country/italy/}{Worldometers.info} reveals the daily new COVID-19 cases in Italy. Italy issued its coronavirus lockdown on March 9th. Figure \ref{fig: italy new cases} seems to elicit a first hypothesis that the effect of lockdown will be seen two weeks after the lockdown (starting March 23rd), and a secondary hypothesis that the effect is seen three weeks after (starting March 30th). 

Motivated by our observation of Italy, we define our primary outcome to be the average (mean) percentage of observed illness during the week of March 30th through April 5th, two weeks after the intervention period. The county-level observed illness data is from the \hyperlink{https://healthweather.us/}{US Health Weather Map by Kinsa Inc.}, which measures illness linked to fever, specifically influenza-like illness. The data is collected and compiled by Kinsa Inc. using more than one million smart thermometers around the country (\citealp{miller2018smartphone}). Although the observed illness data does not directly measure the COVID-19 cases, we believe it reflects an accurate, real-time aspect of the current pandemic. Importantly, unlike outcome measures that are directly related to COVID-19 (e.g., confirmed cases), the observed illness considered in this article is not restricted by the lack of testing for the disease, and it experiences no lags. Similarly, we define our secondary outcome to be the average observed illness during the week of April 6th through April 12th, three weeks after the intervention period. By the time this study protocol is submitted to arxiv, U.S. Health Map has yet to publish the observed illness data starting from April 6th. To deliver the results in a more timely manner, we will analyze the secondary outcome in a later report when observed illness data becomes available at the end of April 12th.


\section{Baseline County-level Characteristics (Observed Covariates)}
\label{sec: confounding vbs}
We use matching to adjust for assorted county-level baseline characteristics. The data of those characteristics come from the United States Census Bureau and the 2020 analytic data from the County Health Rankings and Roadmaps Program, and can be divided into two broad categories: those related to socioeconomic status, and those related to social interaction, health-seeking behaviors, and disease spreading. We list all county-level baseline characteristics adjusted for below:

\begin{itemize}
    \item Population density (population per square miles);
    \item Percentage of female population;
    \item Percentage of population over $65$;
    \item Median household income;
    \item Unemployment rate (percentage of population ages 16 and older unemployed but seeking work);
    \item Percentage graduating from some college (percentage of adults ages 25-44 with some post-secondary education);
    \item Percentage of Hispanic/African American;
    \item Percentage of people in poverty;
    \item Percentage of adult smoking (percentage of adults who are current smokers);
    \item Percentage of flu vaccination (percentage of fee-for-service (FFS) Medicare enrollees that had an annual flu vaccination);
    \item Social associations (number of membership associations per 10,000 population);
    \item Traffic volume (average traffic volume per meter of major roadways in the county);
    \item Rural-Urban Continuum Code (a scheme that distinguishes metropolitan counties by the population size, and nonmetropolitan counties by degree of urbanization and adjacency to a metro area);
    \item Percentage of observed illness during the intervention period (March 16th through March 22nd);
    \item Expected percentage of observed illness during the intervention period (March 16th through March 22nd) predicted by Kinsa Inc.;
    \item Expected percentage of observed illness during the primary outcome period (March 30th through April 5th) prediced by Kinsa Inc..
\end{itemize}
It is important to note that we match on the observed illness at the intervention period, so that we counties in each matched pair have similar disease trajectory prior to or at the point of intervention. After excluding all the counties with any missing baseline characteristics, observed illness data, or the treatment indices (i.e., average social distancing scores from March 16th to March 22nd), there remain 3,031 counties, accounting for about 96.5\% of the total 3,142 counties in the United States. Excluded counties include all counties in Alaska and Hawaii, and a few in the so-called Mountain States.

\section{Analytic Plan}
\subsection{Statistical Matching}
We will perform a matched observational study to compare the primary outcome of counties with high social distancing scores to that of counties with low social distancing scores. Instead of defining treated and control counties based on an ad-hoc cutoff of the continuous social distancing score, we leverage a statistical matching technique known as non-bipartite matching (\citealp{lu2001matching}; \citealp{lu2011optimal}; \citealp{baiocchi2012near}) to directly pair counties based on their continuous social distancing scores. The county with higher social distancing score in each matched pair will be call the \emph{treated} county and the other the \emph{control} county. The collection of all treated counties will be referred to as the \emph{treated group}, and the others the \emph{control group}.

Ideally, we would like to pair a county with a high social distancing score with another county identical in all observed covariate but with a low social distancing score. With moderate number of observed covariates, this task becomes impractical. Instead, we aim to create two groups of counties that are very similar in the mean of each observed covariate. This is known as covariate balance, and we can assess the quality of the match by looking at the standardized difference in the mean of each covariate in the treated and the matched control group. Moreover, we can run a two-sample t-test for each covariate to assess if there is any systematic, statistically significant, difference that is residual after matching. We would aim for a match that has standardized differences of all variables less than $0.1$, and roughly only $1$ out of $20$ $p$-values less than $0.05$.

When covariate balance is deemed good for multiple matches, we prefer one with a more pronounced difference in the social distancing score, possibly at the cost of fewer matched pairs (\citealp{lu2001matching}; \citealp{baiocchi2010building}; \citealp{heng2019instrumental}). In the context of non-bipartite matching, we may strengthen the matched pair difference in the continuous treatment and eliminate some units by adding \emph{sinks} (\citealp{lu2001matching}). To eliminate $e$ counties, $e$ sinks are added to the dataset before matching, with distance between each sink and each county equal to $0$, and the distance between each pair of sinks equal to infinity. We present the final matching results in Section \ref{sec: matching results}.

\subsection{Software}
We will be performing the described statistical matching using the \textsf{R} package \textsf{nbpMatching} (\citealp{lu2011optimal}; \citealp{R_pkg_nbpmatching}) and summarizing the covariate balancing using the \textsf{R} package \textsf{RItools} (\citealp{Hansen2008}; \citealp{R_pkg_RItools}). 

\subsection{Statistical Inference}
Suppose that there are $I$ matched pairs ($I=302$ in our study), and each matched pair contains $2$ units (i.e., counties). Let $d_{ij}$ denote the observed average score of social distancing during March 16th through 22nd of county $j$ in matched pair $i$, $i=1,\dots, I$ and $j=1,2$. For each $i$, let $\overline{d}_{i}=\max \{d_{i1}, d_{i2}\}$ denote the average social distancing score for the treated county in matched pair $i$, and $\underline{d}_{i}=\min \{d_{i1}, d_{i2}\}$ for the control county in matched pair $i$. Let $r_{ij}^{d}$ denote the potential the outcome that we want to study (e.g., can be either the primary or secondary outcome) of county $j$ in matched pair $i$ under the average social distancing score $d$. Following \citet{small2008randomization}, we consider the following proportional treatment effect model on the log scale of the outcome:
\begin{equation*}
    \log r_{ij}^{\overline{d}_{i}}-\log r_{ij}^{\underline{d}_{i}}=\beta (\overline{d}_{i}-\underline{d}_{i}) \quad \text{for some constant $\beta$}.
\end{equation*}
We focus on testing $H_{0}^{\beta_{0}}: \beta=\beta_{0}$ versus $H_{1}^{\beta_{0}}:\beta>\beta_{0}$ for various $\beta_{0}$ (perhaps focusing on $\beta_{0}=0$) and build a 95\% confidence interval for $\beta$. As considered in \citet{small2008randomization}, to incorporate the fruitful information of the observed covariates, we consider testing $H_{0}^{\beta_{0}}$ versus $H_{1}^{\beta_{0}}$ with covariance adjustment. That is, let $R_{ij}$ be the observed outcome of county $j$ in matched pair $i$, we will regress $(\log R_{11}-\beta_{0} d_{11}, \dots,\log R_{I2}-\beta_{0} d_{I2})^{T}$ on the observed covariates and then we will test $H_{0}^{\beta_{0}}$ versus $H_{1}^{\beta_{0}}$ by using the \textsf{senm} command in \textsf{R} package \textsf{sensitivitymult}, which performs the paired Huber's M test in a randomization inference (\citealp{rosenbaum2007sensitivity}), on the residuals from this regression. We will report a 95\% confidence interval for $\beta$ (found by inverting the hypothesis test for $H_{0}^{\beta_{0}}$ under various $\beta_{0}$; see \citealp{rosenbaum2002observational}), a one-sided $p$-value (using the covariance adjustment described above along with the \textsf{senm} command in \textsf{R} package \textsf{sensitivitymult}) for the null hypothesis $H_{0}: \beta=0$ versus $H_{1}:\beta>0$ (i.e., $H_{0}^{\beta_{0}=0}$ versus $H_{1}^{\beta_{0}=0}$). If the $p$-value is $<0.05$, we will consider conducting a sensitivity analysis using the method in \citet{rosenbaum1989sensitivity} and report the ``worst-case" one-sided $p$-value under each prespecified sensitivity parameter (i.e., the largest one-sided $p$-value over all possible arrangements of the hypothesized unmeasured counfounders given each sensitivity parameter). We would then report the sensitivity value, the largest sensitivity parameter at which the ``worst-case" $p$-value is still $\leq 0.05$ \citep{rosenbaum2002observational, zhao2019sensitivity}.

\section{Matching Results}
\label{sec: matching results}

\begin{figure}[t]
    \centering
    \includegraphics[scale = 0.7]{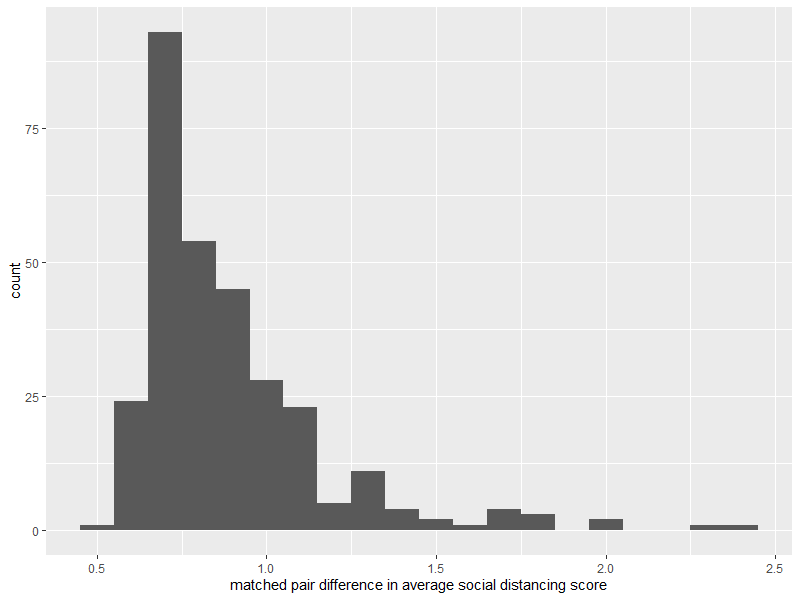}
    \caption{Distribution of treated-minus-control difference in the social distancing score after optimal non-bipartite matching}
    \label{fig: hist of matched pair diff in sd score}
\end{figure}

Table \ref{tbl: balance table near far} summarizes the covariate balance and average social distancing scores of the treated and control groups. We form a total of $302$ pairs of two counties using optimal non-bipartite matching, which roughly corresponds to using $40\%$ of all counties in the US. We observe no statistically significant difference among the treated and control groups in any of the $16$ observed covariates adjusted for. The average social distancing score of the treated group is $1.17$, corresponding to a letter grade slightly better than $D$ (which translates to more than $40\%$ decrease in the distance traveled and more than $60\%$ decrease in the visits to non-essential venues.). On the other hand, the average score is $0.29$ for the control group, corresponding to a letter grad slightly better than $F$. Figure \ref{fig: hist of matched pair diff in sd score} further plots the distribution of the treated-minus-control difference of the social distancing scores. The median treated-minus-control difference in the social distancing scores is $0.80$. The match and the balance table are available via author's Github page: \url{https://github.com/bzhangupenn/Social_Distancing_Match}.

\begin{table}[ht]
\small
\centering
\caption{Covariate balance after matching. $302$ pairs of two counties are formed. After matching, the magnitudes of all standardized differences are less than $0.1$, and all $p$-values larger than $0.1$, suggesting no systematic difference between the encouraged and control groups. The average social distancing score of the treated group is $1.17$, corresponding to more than $40\%$ reduction in travel. The average social distancing score of the control group is $0.29$, roughly corresponding to less than $25\%$ reduction in travel.}
\begin{tabular}{lcccc}
\hline
& Control & Treated & Std.diff. & $p$-value\\ \hline \\
Average social distancing score &  0.29   &1.17    &2.71   & 0.00 \\ 
 Number of counties & 302   & 302    &   &      \\
\textbf{Covariates} &           &          &          \\ 
\hspace{0.5 cm}\% Female & 0.50     & 0.50    & -0.01 & 0.89     \\
\hspace{0.5 cm}\% Above 65  & 0.20    & 0.21    & 0.09 & 0.12     \\ 
\hspace{0.5 cm}\% Unemployment   & 0.04   & 0.04  & 0.00 & 0.99   \\
\hspace{0.5 cm}\% Smoking &0.17 &0.17 &-0.07 &0.26 \\
\hspace{0.5 cm}\% Flu vaccination &0.41 &0.41 &0.04 &0.46 \\
\hspace{0.5 cm}\% Some college &0.61 &0.62 &0.06 &0.29 \\
\hspace{0.5 cm}\% Black/Hispanic  &0.10   & 0.10  & -0.02 &0.73   \\
\hspace{0.5 cm}\% Poverty &13.67 &13.87 &0.04 &0.52 \\
\hspace{0.5 cm}Median household income  & 52990.65   & 52855.70 & -0.01 &0.83       \\ 
\hspace{0.5 cm}Social association &13.93 &13.56 &-0.06 &0.31 \\
\hspace{0.5 cm}Traffic &72.07 &78.90 &0.07 &0.26 \\
\hspace{0.5 cm}Rural-Urban Continuum Code &5.51 &5.73 &0.08 &0.16 \\
\hspace{0.5 cm}Population density  &85.52  &90.00  &0.03 &0.65 \\
\hspace{0.5 cm}\% Obs. illness (Mar 16th - Mar 22nd) &4.37 &4.35 &-0.03 &0.61 \\
\hspace{0.5 cm}\% Exp. illness (Mar 16th - Mar 22nd) &4.11  &4.14   &0.03 &0.62\\
\hspace{0.5 cm}\% Exp. illness (Mar 30th - Apr 5th) &3.03  &3.11   &0.09 &0.13\\
\hline
\end{tabular}
\label{tbl: balance table near far}
\end{table}

\clearpage
\bibliography{coronavirus_refs}


\end{document}